# Orbital torque switching in perpendicularly magnetized materials

Yuhe Yang[1,†], Ping Wang[2,8,†], Jiali Chen[3,4,†], Delin Zhang[1,2,8,†,*], Chang Pan[5], Shuai Hu[2], Ting Wang[1], Wensi Yue[2], Cheng Chen[2], Wei Jiang[3,4,*], Lujun Zhu[6], Xuepeng Qiu[5], Yugui Yao[3,4], Yue Li[7], Wenhong Wang[1,2,7], Yong Jiang[1,2,*]

1. School of Material Science and Engineering, Tiangong University, Tianjin, 300387, China.
2. Institute of Quantum Materials and Devices, School of Electronic and Information Engineering, Tiangong University, Tianjin, 300387, China.
3. Centre for Quantum Physics, Key Laboratory of Advanced Optoelectronic Quantum Architecture and Measurement (MOE), School of Physics, Beijing Institute of Technology, Beijing, 100081, China.
4. Beijing Key Lab of Nanophotonics & Ultrafine Optoelectronic Systems, School of Physics, Beijing Institute of Technology, Beijing, 100081, China.
5. Shanghai Key Laboratory of Special Artificial Microstructure Materials and Technology and School of Physics Science and Engineering, Tongji University, Shanghai 200092, China.
6. School of Physics and Information Technology, Shaanxi Normal University, Xi'an 710062, China.
7. School of Physical Science & Technology, Tiangong University, Tianjin, 300387, China.
8. Cangzhou Institute of Tiangong University, Cangzhou, 061000, China.

†These authors contributed equally: Yuhe Yang, Ping Wang, Jiali Chen, Delin Zhang
*Corresponding authors. Email: zhangdelin@tiangong.edu.cn (D.L.Z.), wjiang@bit.edu.cn (W.J.) and yjiang@tiangong.edu.cn (Y.J.)

## Abstract

The orbital Hall effect in light materials has attracted considerable attention for developing novel orbitronic devices. Here we investigate the orbital torque efficiency and demonstrate the switching of the perpendicularly magnetized materials through the orbital Hall material (OHM), i.e., Zirconium (Zr). The orbital torque efficiency of approximately 0.78 is achieved in the Zr OHM with the perpendicularly magnetized [Co/Pt]$_3$ sample, which significantly surpasses that of the perpendicularly magnetized CoFeB/Gd/CoFeB sample (approximately 0.04). Such notable difference is attributed to the different spin-orbit correlation strength between the [Co/Pt]$_3$ sample and the CoFeB/Gd/CoFeB sample, which has been confirmed through the theoretical calculations. Furthermore, the full magnetization switching of the [Co/Pt]$_3$ sample with a switching current density of approximately $2.6 \times 10^6$ A/cm$^2$ has been realized through Zr, which even outperforms that of the W spin Hall material. Our finding provides a guideline to understand orbital torque efficiency and paves the way to develop energy-efficient orbitronic devices.

The demand for energy-efficient electronic devices has accelerated the efforts to carry and store information by utilizing the spin angular momentum (**S**) of electrons over the past decade. Spin Hall effect (SHE) is a physical phenomenon in which the transverse spin currents can be generated under an external electric field (**E**) in materials with strong spin-orbit coupling (SOC) [1-5]. The spin current can induce spin-orbit torque (SOT) and switch the magnetization of the adjacent ferromagnetic (FM) layer for designing energy-efficient spintronic memory and logic devices [3-8]. Most recent efforts have shifted towards the exploration of the orbital angular momentum (**L**) in materials without depending on SOC for developing orbitronic memory and logic devices [9-21]. Generally, orbital quenching caused by crystal fields prevents significant dynamical behaviors of orbital degrees. However, recent studies have shown that the orbital texture caused by orbital hybridization can generate finite orbital angular momentum along the direction of **E**×**k** when the **E** is applied [11,12]. That leads to another physical phenomenon of orbital Hall effect (OHE), where the orbital current can be generated under the **E**. The large orbital Hall conductivity ($\sigma_{OHE}$) of the light materials (LMs) with weak SOC can efficiently convert charge current to orbital current [12,15]. Furthermore, the orbital current can induce orbital torque (OT) for magnetization switching, by entering the adjacent FM materials and converting them into spin current via the SOC of the FM materials [13-16]. The orbital-torque efficiency ($\xi_{OT}$) depends on not only the orbital Hall angle ($\theta_{OHE}$) of the LMs but also the orbital-to-spin conversion coefficient ($\eta_{L-S}$) of the adjacent FM layers [15,16].

Recently, the $\xi_{OT}$ and $\theta_{OHE}$ have been experimentally investigated in the LMs through the methods like magneto-optical Kerr effect, spin-torque ferromagnetic resonance, harmonic Hall effect, anomalous Hall effect (AHE), and Hanle magnetoresistance [15,16,22-33]. The $\xi_{OT}$ or $\theta_{OHE}$ of Ti [15,23], Cr [24-26,28], and Mn [29] systems have been evaluated to be approximately 0.13, 0.28, and 0.016, respectively. Most recently, the experiment has confirmed that the contribution originated from OHE dominates the torque compared to the SHE in the light metal Zr systems, which has the large orbital Hall conductivity ($\sigma_{OHE}$) of 5300 $\hbar/e$ $\Omega^{-1}$ cm$^{-1}$ and small spin Hall conductivity ($\sigma_{SHE}$) of -170 $\hbar/e$ $\Omega^{-1}$ cm$^{-1}$ [22,27]. To date, however, there has been limited exploration of the experimental magnetization switching of the FM layer with perpendicular magnetic anisotropy (PMA) through the OT. This is primarily due to the demanding requirements of both PMA and large $\eta_{L-S}$ for the adjacent FM layer, let alone the industrial-compatible PMA samples for designing the energy-efficient orbitronic memory and logic devices.

In this work, we conducted experimental investigations into the $\xi_{OT}$ and magnetization switching via the OT of the orbital Hall material (OHM), Zr, with different adjacent PMA FM materials, (i.e., [Co (0.3 nm)/Pt (0.7 nm)]$_3$ sample, hereafter termed [Co/Pt]$_3$ and Co$_{20}$Fe$_{60}$B$_{20}$ (0.8 nm)/Gd (1.2 nm)/Co$_{20}$Fe$_{60}$B$_{20}$ (1.1 nm) sample, hereafter termed CFB/Gd/CFB). We found that the Zr OHM can easily switch the [Co/Pt]$_3$ sample with lower switching current density ($J_s$) than the CFB/Gd/CFB sample. The observed distinction is also manifested by their different $\xi_{OT}$, with the Zr/[Co/Pt]$_3$ sample having larger $\xi_{OT}$ (~ 0.78) compared to the Zr/CFB/Gd/CFB sample (~ 0.04), as characterized through the second harmonic measurement.

Meanwhile, first-principles calculations have been employed to calculate the spin-orbit correlation [which is proportional to the $\eta_{L-S}$] of the [Co/Pt]$_3$ and CFB/Gd/CFB samples, which indicates a larger $\eta_{L-S}$ of the [Co/Pt]$_3$ than that of the CFB/Gd/CFB samples, providing insight into the underlying physical mechanism for experimental results. These results highlight the significance of the spin-orbit correlation of the PMA FM layer, offering valuable insights for the development of novel orbitronic electronics.

**Spin-orbit torque (SOT) and orbital torque (OT)**

In the SOT system, the pivotal factor lies in the properties of the spin Hall Materials (SHMs). In the SHM/FM heterostructures, the conversion of charge current into spin current occurs through the SHMs due to their SOC. The SOT efficiency ($\xi_{SOT}$) is determined by the spin Hall angle ($\theta_{SHE}$), as described by $\theta_{SHE} = (2e/\hbar) \sigma_{SHE}/\sigma_{xx}$, depends critically on the $\sigma_{SHE}$ and thus the SOC of SHMs, which barely relies on the properties of the adjacent FM layers [3-8], as shown in Fig. 1a. Common SHMs encompass heavy metals (HMs, e.g. Ta, W, Pt), topological materials (TMs, e.g. Bi$_{0.9}$Sb$_{0.1}$, Bi$_2$Se$_3$, WTe$_2$) [3-8], etc. Generally, the HMs have the $\sigma_{SHE}$ around $10^2 \sim 10^3$ $\hbar/e$ $\Omega^{-1}$ cm$^{-1}$ and the $\theta_{SHE}$ around 0.1-0.4 [2]. The TMs with the strong SOC could have even larger $\sigma_{SHE} \sim 10^4$ $\hbar/e$ $\Omega^{-1}$ cm$^{-1}$, thus the $\theta_{SHE}$ can reach 2-75 [3-8]. For the OT system, as depicted in Fig. 1b, the properties of both the OHM and the adjacent FM layers play an important role. In the OHM/FM heterostructures, the charge current can be converted to orbital current in the LMs, then transmitted into the FM layers and converted to spin current through the orbital-to-spin conversion of the FM layers [15,16].

The $\xi_{OT}$ depends on the $\eta_{L-S} \cdot \theta_{OHE}$, as $\eta_{L-S} \cdot \theta_{OHE} = (2e/\hbar)\eta_{L-S} \cdot \sigma_{OHE}/\sigma_{xx}$ [15,16,23,24]. The LMs, such as Ti, Zr, Cr, Mn, Nb, Ru, and Cu, exhibit the $\sigma_{OHE}$ of about $10^3$ $\hbar/e$ $\Omega^{-1}$ cm$^{-1}$, even surpassing the $\sigma_{SHE}$ of the HMs [10,11,15,27]. On the other hand, by far, the FM materials employed to investigate the $\xi_{OT}$ of the OHM/FM systems, only focus on those with the relatively larger $\eta_{L-S}$ (~ 0.0192-0.0455), e.g. CoFe, Co, Ni, which, however, are lack of PMA materials [16]. To realize the efficient magnetization switching for realistic industrial applications through the OHE, exploring the novel OHM/FM systems with the larger $\sigma_{OHE}$ of the OHMs and the larger $\eta_{L-S}$ of the PMA FM layers will be very crucial.

**Orbital Torque Efficiency**

To experimentally investigate the $\xi_{OT}$ as well as the influence from the PMA FM layer, we designed the samples with the Zr OHM and two kinds of PMA FM layers, [Co/Pt]$_3$ and CFB/Gd/CFB PMA samples. The samples, Zr (10.0 nm)/[Co/Pt]$_3$ and Zr (10.0 nm)/CFB/Gd/CFB, as well as the W SHM reference samples with the identical samples, were deposited on the Al$_2$O$_3$ (0001) single crystal substrates and the thermal-oxide Si/SiO$_2$(300 nm) substrates, respectively (see the sample preparation of the METHODS). To characterize the crystal structure of the Zr layer, the 50-nm-thick Zr thin films were prepared with the same experimental condition as the Zr/FM heterostructures and then measured by X-ray diffraction. The polycrystalline structure with the (001) orientation is observed with the (002) and (004) peaks, as shown in Supplementary Figs. 1a and 1b (see Supplementary Note 1). Meanwhile, the surface morphology of the Al$_2$O$_3$/Zr (10.0 nm)/[Co (0.3 nm)/Pt (0.7 nm)]$_3$ and Si/SiO$_2$/Zr (10.0 nm)/CFB (0.8 nm)/Gd (1.2 nm)/CFB (1.1 nm) samples were characterized by atomic force microscope (AFM). We found that the Zr (10.0 nm)/[Co/Pt]$_3$ and Zr (10.0 nm)/CFB/Gd/CFB samples show a very smooth surface with an average surface roughness ~ 0.139 nm and 0.144 nm, respectively.

Subsequently, the torque efficiency was determined by the harmonic Hall voltage measurement, which can quantitatively characterize the effective magnetic field originating from the SHE or OHE induced by the alternating current (a.c.) [34]. Figures 2a, 2d, and Supplementary Figs. 2a, 2d show the schematic of the harmonic Hall voltage measurement for the Zr (10.0 nm)/[Co/Pt]$_3$ and Zr (10.0 nm)/CFB/Gd/CFB devices, as well as W (5.0 nm)/[Co/Pt]$_3$ and W (5.0 nm)/CFB/Gd/CFB devices. In these configurations, the a.c. current is applied along the $x$-axis with sweeping the in-plane external magnetic field ($H_{ext}$) along the $x$-axis and $y$-axis, corresponding to longitudinal and transverse schemes, enabling the separate characterization of the longitudinal effective magnetic field ($H_L$) and transverse effective magnetic field ($H_T$) resulting from the current-induced torque. The samples were patterned into Hall bars (140 μm × 16 μm), and harmonic measurements were conducted. The measured results are presented in Fig. 2 (longitudinal scheme) and Supplementary Fig. 2 (transverse scheme), the $H_L$ and $H_T$ can be obtained from the equations below:

$$H_L = -2\frac{B_x \pm 2\xi B_y}{1-4\xi^2} \quad \text{and} \quad H_T = -2\frac{B_y \pm 2\xi B_x}{1-4\xi^2}$$ [34,35], where $B_x \equiv \frac{\partial V_{2\omega,L}}{\partial H} \Big/ \frac{\partial^2 V_{\omega,L}}{\partial H^2}$ and $B_y \equiv \frac{\partial V_{2\omega,T}}{\partial H} \Big/ \frac{\partial^2 V_{\omega,T}}{\partial H^2}$, $\xi$ is the ratio between the planar Hall resistance ($R_{PHE}$) and the anomalous Hall resistance ($R_{AHE}$) from the contributions from planar Hall effect (PHE) and anomalous Hall effect (AHE), respectively. The symbol ± represents the

magnetization direction, oriented either along +z or –z. We find that the first harmonic voltage ($V_\omega$) with $H_{ext}$ can be well fitted by the parabolic function, suggesting that the samples are magnetically saturated and undergo single-domain motion during sweeping of the $H_{ext}$. The second harmonic voltage ($V_{2\omega}$) exhibits a linear variation in response to $H_{ext}$, as shown in Figs. 2b, 2c, 2e, 2f, and Supplementary Figs. 2b, 2c, 2e, and 2f. The consistent sign of the slope in the longitudinal scheme for positive and negative magnetized states, along with the inverse sign of the slope in the transverse scheme, suggests the presence of the torque originated from the OHE of Zr and SHE of W. The damping-like torque efficiency ($\theta_{DL}$) can be extracted using the equation of

$$\theta_{DL} = \frac{2e}{\hbar} M_S t_F \frac{\mu_0 H_L}{J_e}$$

(see Supplementary Note 2) [34,35]. For estimating the $\theta_{DL}$, we focus on determining the effected current density $J_e$ in the Zr OHM and W SHM without considering the contribution of current-induced SOT in the single [Co/Pt]$_3$ and CFB/Gd/CFB FM layers. The $\xi_{OT,DL}$ is estimated to be ~ 0.78 for the Zr (10.0 nm)/[Co/Pt]$_3$ sample but is only $\xi_{OT,DL}$ ~ 0.04 for the Zr (10.0 nm)/CFB/Gd/CFB sample. This discrepancy suggests that the $\eta_{L-S}$ of the FM layer plays a crucial role in the OHE. The fitted $\xi_{SOT,DL}$ is ~ 0.53 and ~ 0.34 for the reference W (5.0 nm)/[Co/Pt]$_3$ and W (5.0 nm)/CFB/Gd/CFB samples, respectively, which is consistent with the reported values [36,37]. Additionally, the sign of $\xi_{OT,DL}$ of the Zr (10.0 nm)/[Co/Pt]$_3$ sample is consistent with the W (5.0 nm)/[Co/Pt]$_3$ sample because the positive sign of $\sigma_{OHE}$ of Zr and negative $\eta_{L-S}$ of [Co/Pt]$_3$ should lead to the negative $\xi_{OT}$ of Zr (10.0 nm)/[Co/Pt]$_3$ [38,39], consistent with the negative $\theta_{SHE}$ of W [36,37]. We also summarize the $\xi_{OT}$ as a function of the $\sigma_{OHE}$ for the OHMs with different FM layers [15,22-27,30-33], as plotted in Fig. 2g. To the best of our knowledge, the value of the $\xi_{OT}$ of Zr (10.0 nm)/[Co/Pt]$_3$ in our work is the largest one so far, suggesting that the [Co/Pt]$_3$ PMA layer has a large $\eta_{L-S}$ and can be selected as a dominant FM layer for conversion of orbital current to spin current.

**Orbital Torque Switching**

To further investigate the efficiency of the magnetization switching through the Zr OHM and W SHM, we prepared the Zr (10.0 nm)/[Co/Pt]$_3$ sample and a reference sample with the structure of W (5.0 nm)/[Co/Pt]$_3$. Figures 3a and 3c show the $R_{AHE}$ as functions of the out-of-plane $H_{ext}$ for the [Co/Pt]$_3$ sample deposited on the Zr OHM and W SHM, respectively. The square $R_{AHE}$ vs. $H_{ext}$ loops indicate their excellent PMA properties. Magnetometry was also utilized to confirm the PMA of the [Co/Pt]$_3$ sample (see Supplementary Note 3 and Supplementary Fig. 3). Subsequently, the samples were patterned into the Hall bar devices, and the current-induced orbital/spin torque magnetization switching was performed under an in-plane $H_{ext}$ along $x$-axis direction, ranging from ±100 Oe to ±500 Oe with the 20-ms write direct-current (d.c.) pulse width. Figure 3b and Supplementary Fig. 4a illustrate the magnetization switching of the Zr (10.0 nm)/[Co/Pt]$_3$ sample. Complete magnetization switching is observed with the assistance of the positive and negative in-plane $H_{ext}$ at $J_s \sim 2.6 \times 10^6$ A/cm$^2$. Furthermore, upon reversing the applied in-plane field direction, the polarity of the current-induced switching loop also reverses, excluding the thermal effect. Since the $R_{AHE}$ values obtained by the field sweeping and current sweeping are similar, we conclude that the full magnetization switching of the [Co/Pt]$_3$ sample has been achieved. However, for the W (10.0 nm)/[Co/Pt]$_3$ sample, as shown in Fig. 3d and Supplementary Fig. 4b, only partial magnetization switching occurs with the assistance of the positive and negative in-plane $H_{ext}$ at $J_s \sim 6.6 \times 10^6$ A/cm$^2$ compared to the $R_{AHE}$ values obtained by field sweeping. Complete magnetization switching cannot be observed even at high $J_s$. It is noteworthy that the $J_s$ of Zr (10.0 nm)/[Co/Pt]$_3$ ($J_s \sim 2.6 \times 10^6$ A/cm$^2$) is comparable with or better than that of W (5.0 nm)/[Co/Pt]$_3$ ($J_s \sim 6.6 \times 10^6$ A/cm$^2$), suggesting that the [Co/Pt]$_3$ sample with large $\eta_{L-S}$ can more efficiently convert orbital current to spin current through the Zr OHM.

Subsequently, the other PMA CFB/Gd/CFB sample is utilized to study the current-induced orbital/spin-torque magnetization switching through the Zr OHM and W SHM under the same measured condition. Figures 3e, 3g, and Supplementary Figs. 5a and 5b display the square $R_{AHE}$ vs. $H_{ext}$ and $M/M_s$ vs. $H_{ext}$ loops of the Zr (10.0 nm)/CFB/Gd/CFB and W (5.0 nm)/CFB/Gd/CFB samples, respectively, confirming their good PMA properties. Figure 3f and Supplementary Fig. 6a show the magnetization switching of the Zr (10.0 nm)/CFB/Gd/CFB sample with the assistance of both the positive and negative in-plane $H_{ext}$ at $J_s \sim 4.9 \times 10^6$ A/cm$^2$. We note that it is a partial magnetization switching, the possible reason could be the relatively small $\xi_{OT,DL} \sim 0.04$. However, for the W (5.0 nm)/CFB/Gd/CFB sample as shown in Fig. 3h and Supplementary Fig. 6b, the full magnetization switching occurs with the assistance of both the positive and negative in-plane $H_{ext}$ at $J_s \sim 1.6 \times 10^6$ A/cm$^2$. Due to the relatively larger $\theta_{SHE}$, the $J_s$ is lower than that of the Ta (5.0 nm)/CFB/Gd/CFB sample with $J_s \sim 2.0 \times 10^7$ A/cm in our previous work [4]. Although the Zr layers deposited on the Al$_2$O$_3$ and Si/SiO$_2$ substrates show the almost same crystalline qualities, the capability of the magnetization switching is different. The main reason will be the contribution of the PMA FM layers, [Co/Pt]$_3$ and CFB/Gd/CFB.

**Spin-orbit Correlation Function calculation**

One of the critical factors influencing the torque efficiency of OHE is the $\eta_{\text{L-S}}$ of the FM materials, as indicated by the aforementioned harmonic results and orbital torque switching, which describes the orbital-to-spin current conversion efficiency. To gain a deeper understanding of the intrinsic relationship between the torque efficiency, OT switching, and the $\eta_{\text{L-S}}$ of the FM layer in the OHE system, we designed the CoPt$_2$ and Co$_2$Fe$_6$ structures to represent the experimental [Co/Pt]$_3$ and CFB/Gd/CFB samples, respectively, and conducted full electronic simulations to calculate their spin-orbit correlation function $\langle \boldsymbol{L}\cdot\boldsymbol{S}\rangle$ based on density functional theory (see methods for calculation details). Figures 4a and 4b show the side views for CoPt$_2$ and Co$_2$Fe$_6$ structures, respectively, with similar thicknesses as the experiments. The magnitude of $\eta_{\text{L-S}}$ is roughly proportional to the $\langle \boldsymbol{L}\cdot\boldsymbol{S}\rangle$ [16]. The band-resolved spin-orbit correlation function $\langle \boldsymbol{L}\cdot\boldsymbol{S}\rangle_{nk}$ was calculated after constructing the effective Hamiltonian using the Wannier90 package (see Supplementary Note 4, Supplementary Figs. 7-10). The results are summarized in Fig. 4, where red and blue colored areas denote strong positive and negative correlations, respectively. In Fig. 4a for the CoPt$_2$ structure, multiple spin-orbit correlation hotspots below the Fermi level ($E_F$) (e.g., along the Γ-X and Γ-M line) lead to a large $\eta_{\text{L-S}}$, indicating the strong orbital-to-spin conversion efficiency. In addition, the magnetic moments are primarily located on the Co atoms (~ 2.1 $\mu_B$) with small contributions from the Pt atoms (~ 0.3 $\mu_B$). The magnetic moment of Pt derives from strong proximity effect between the Co and Pt, resulting in the substantial exchange splitting of Pt (projected by band structure in Supplementary Fig. 11). By checking the wave function, we find that correlation $\langle \mathbf{L} \cdot \mathbf{S} \rangle_{nk}$ with including the SOC is contributed by $d$ states of Co and $d$ states of Pt in the CoPt$_2$ structure, as shown in Supplementary Fig. 11a. Atom-resolved band structure analysis further supports this, showing that the majority of the spin-orbit correlation hotspots are mainly contributed by the Pt bands (see Supplementary Fig. S12). On the contrary, for the Co$_2$Fe$_6$ structure, we observe limited spin-orbit correlation hotspots below the $E_F$, indicating much weaker correlation strength compared to the CoPt$_2$ structure, as shown in Fig. 4b. By checking the wave function, we can see that the $d$ states of Co and Fe in the Co$_2$Fe$_6$ structure, especially the majority atom Fe, play a leading role in contributing to correlation $\langle \mathbf{L} \cdot \mathbf{S} \rangle_{nk}$ with including the SOC, as shown in Supplementary Fig. 11b. These results align with our experimental results, revealing a relatively high $\xi_{OT}$ of the Zr/CoPt$_2$ structure compared to the Zr/Co$_2$Fe$_6$ structure. This underscores the significance of the $\eta_{L-S}$ of the FM materials in the OHE devices, which provides another pathway for improving the orbital torque efficiency.

**Discussion**

To summarize, we have investigated the $\xi_{OT}$ and magnetization switching of the PMA materials (i.e. [Co/Pt]$_3$ and CFB/Gd/CFB) through the OT of the Zr OHM. The largest $\xi_{OT}$ ~ 0.78 was obtained in Zr/[Co/Pt]$_3$ due to both the large OHE of Zr and the strong spin-orbit correlation function of the [Co/Pt]$_3$ strucure. Furthermore, the switching of the PMA materials with the large $\eta_{\text{L-S}}$ via the Zr OHM ($J_s \sim 2.6 \times 10^6$ A/cm$^2$) is much more efficient than that of the PMA materials with the small $\eta_{\text{L-S}}$ via the Zr OHM ($J_s \sim 4.9 \times 10^6$ A/cm$^2$). These experimental results have also been confirmed through the spin-orbit correlation function calculation, where the CoPt$_2$ structure possesses stronger spin-orbit correlation function strength than that of the Co$_2$Fe$_6$ structure. Our results highlight the crucial relationship between the $\xi_{\text{OT}}$ and the spin-orbit correlation function of the PMA materials and establish a direct link between the magnetization switching and the properties of the PMA FM layer.

Our experimental and calculated results carry important implications for developing efficient orbitronic memory and logic devices. Previous studies about SOT devices have only emphasized the exploration of strong SOC SHMs, while recent investigations about the orbital torque devices have focused on the $\xi_{\text{OT}}$ values of the OHMs. Our findings not only demonstrate the OT switching of the PMA materials but also elucidate the physical mechanism to realize the efficient OT switching by engineering the PMA materials with large $\eta_{\text{L-S}}$. Moreover, we note that the PMA materials with large $\eta_{\text{L-S}}$ enable to enhancement of the $\xi_{\text{OT}}$ and switching efficiency, which, however, could be further optimized, such as the damping constant. Normally, the strong spin-orbit coupling of the FM layer will contribute to the relatively larger damping constant. Therefore, we need to engineer the PMA materials (e.g. FePd) which have both large $\eta_{\text{L-S}}$ and low damping constant, or synthetic antiferromagnetic structures in which one PMA layer with large $\eta_{\text{L-S}}$ (e.g. [Co/Pd]$_n$ or [Co/Pt]$_n$) and the other PMA layer with small $\eta_{L\text{-}S}$ and low damping constant (e.g. CoFeB). Consequently, our study offers a new design guideline for novel memory and logic device optimization that may overcome the difficulties of SOT devices and explore industry-compatible orbitronic devices with excellent performance.

## METHODS

**Sample Preparation and Characterization**

All the samples were prepared by utilizing an ultrahigh vacuum magnetron sputtering system with a base pressure lower than $5.0 \times 10^{-8}$ Torr. The Zr (50.0 nm), Zr (10.0 nm)/[Co (0.3 nm)/Pt (0.7 nm)]$_3$ and W (5.0 nm)/[Co (0.3 nm)/Pt (0.7 nm)]$_3$ samples were deposited on the Al$_2$O$_3$ (0001) single crystal substrates, and the Zr (50.0 nm), Zr (10.0 nm)/CFB (0.8 nm)/Gd (1.2 nm)/CFB (1.1 nm)/MgO (2.0 nm)/W (2.0 nm), and W (5.0 nm)/CFB (0.8 nm)/Gd (1.2 nm)/CFB (1.1 nm)/MgO (2.0 nm)/W (2.0 nm) samples were deposited on the Si/SiO$_2$ substrates. During the deposition process, the Ar pressure was 3 mTorr. The Zr layer on the Al$_2$O$_3$ substrate was prepared with the substrate heating at 300 °C and the Zr layer on the Si/SiO$_2$ substrate was prepared with the post-annealing at 300 °C. The crystal structure and quality of the Zr (50.0 nm) thin films on Al$_2$O$_3$ and Si/SiO$_2$ substrates are characterized by X-ray diffraction (XRD) and atomic force microscope (AFM). The *M-H* curves were measured by a vibrating sample magnetometer (VSM).

**Device Patterning and Testing**

The samples were patterned into Hall bar devices by using standard photolithography and an Ar ion milling technique. Ti (10.0 nm)/Pt (100.0 nm) were deposited as the contact electrodes, which were made by d.c. sputtering and lift-off technology. For the harmonic Hall voltage measurements, the devices were first magnetized to a saturation state by applying a large magnetic field (± 1.6 T) perpendicular to the film plane. A sinusoidal a.c. current with an amplitude of 8 mA or 16 mA and frequency of 13.7 Hz was applied by a Keithley 6221 source meter, and the first harmonic Hall voltages ($V_\omega$) and second harmonic Hall voltages ($V_{2\omega}$) are simultaneously measured through two Stanford SR830 lock-in amplifiers while sweeping $H_{ext}$ along or transversely to the current direction. The $R_{AHE}$ - $H$ curves were obtained by using a Keithley 2400 source meter. The $R_{AHE}$ - $J_s$ curves were detected by using Keithley 6221 and 2182A with a fixed magnetic field of $\pm 100$ Oe ~ $\pm 500$ Oe along the current direction. The current density above was calculated by considering the shunting effect of the FM layer in all devices. All the measurements were performed at room temperature.

**Theoretical calculations**

$CoPt_2$ and $Co_2Fe_6$ structures along the [001] direction were simulated using the face-centered cubic and body-centered cubic phases, respectively. We performed first-principles calculations based on the density functional theory (DFT) as implemented in the Vienna *ab* initio simulation package (VASP) [40,41], which is treated by the projector-augmented plane-wave (PAW) method and utilizes a plane wave basis set [42]. The exchange-correlation potential terms were considered at the level of generalized gradient approximation (GGA) within the scheme of Perdew-Burke-Ernzerhof (PBE) functional [43]. The plane-wave cutoff energy is chosen as 450 eV, and we sample the Brillouin zone on $9\times9\times1$ and $12\times12\times1$ regular mesh for the self-consistent calculations of $CoPt_2$ and $Co_2Fe_6$, respectively. The geometric optimizations were performed without any constraint with a convergence criterion of $10^{-6}$ eV. To avoid the interaction between the layers along the *z* direction, we introduced a vacuum layer with a thickness of 15 Å.

To investigate the spin-orbit correlation $\langle \boldsymbol{L}\cdot\boldsymbol{S} \rangle$ for $Co_2Fe_6$ and $CoPt_2$, which gives the orbital to spin conversion for the orbital Hall current (i.e., $\sigma_{SH} \sim \langle \boldsymbol{L}\cdot\boldsymbol{S} \rangle \sigma_{OH}$ [10,44], the Kohn-Sham wave functions were projected onto highly symmetric atomic orbitals like Wannier functions (specifically, Co-*d* and Pt-*s*, *p*, *d* orbitals for $CoPt_2$, Co-*s*, *p*, *d* and Fe-*s*, *p*, *d* orbitals for $Co_2Fe_6$) and we constructed tight-binding Hamiltonians utilizing the WANNIER90 package [45,46]. The obtained Hamiltonian is further applied to obtain the spin-orbit correlation $\langle \boldsymbol{L}\cdot\boldsymbol{S} \rangle_{nk}$ with the crystal momentum *k* in the band $n$ by computing $\langle nk | \boldsymbol{L}\cdot\boldsymbol{S} | nk \rangle$. $\boldsymbol{L}$ refers to orbital angular momentum, which is treated as an atomic angular momentum operator. $\boldsymbol{S}$ is the spin angular momentum operator, and the direct product of these two operators will yield the $\boldsymbol{L}\cdot\boldsymbol{S}$ matrix.

**Acknowledgments:** This work was supported by the National Key R&D Program of China (2022YFA1204003) and the National Natural Science Foundation of China (Grant Nos. 52271240, U23A20551, 12204037, 52061135205, 51971023, 51971024, 51927802, 52271186, 52201292). D.L.Z gratefully acknowledges the research funding provided by the Cangzhou Institute of Tiangong University (Grant No. TGCYY-F-0201). P.W. gratefully acknowledges this work was supported by Hebei Natural Science Foundation (Grant No. E2023110012).

**Author contributions:** Y.H.Y., P.W., J.L.C., and D.L.Z. contributed equally to this work. D.L.Z initialized and conceived this work. Y.J. coordinated and supervised the project. D.L.Z., Y.H.Y., and P.W. conceived the experiments and designed all the samples. Y.H.Y. prepared all the samples, T.W., W.S.Y., C.C., and Y.H.Y patterned the Hall bar devices and carried out the current-induced orbital/spin-torque magnetization switching experiments. C.P., S.H., and X.P.Q. carried out the second harmonic measurements. J.L.C., W.J., and Y.G.Y. performed the first-principles calculations. L.J.Z. conducted the crystalline characterizations. Y.L., and W.H.W. help analyze the experimental data. D.L.Z., Y.H.Y., and P.W. wrote the manuscript. All the authors discussed the results and commented on the manuscript.

**Competing interests:** Authors declare no competing interests.

**Data and materials availability:** All data are available in the manuscript or the Supplementary Materials.

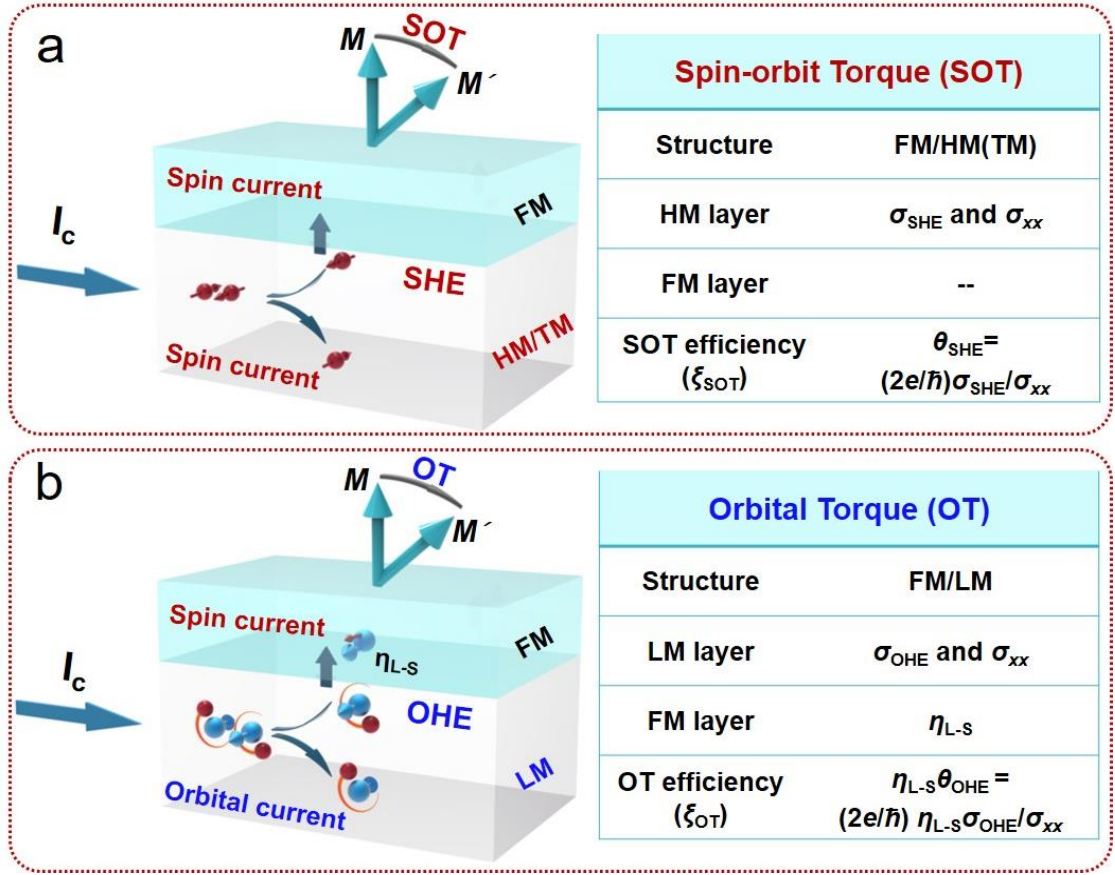

**Fig. 1. Spin-orbit torque (SOT) and orbital torque (OT). a.** Schematic of the SOT system. Spin current generated from the heavy metals (HMs) or topological materials (TMs) due to the spin Hall effect is injected into the adjacent ferromagnetic (FM) layer, exerting the SOT. The torque efficiency ($\xi_{SOT}$) of SOT is dominated by the spin Hall angle ($\theta_{SHE}$) of the HMs or TMs. **b.** Schematic of the OT system. Orbital current originated from the light materials (LMs) due to the orbital Hall effect is injected into the adjacent FM layer, then the orbital current is converted to spin current in the FM layer due to the orbital-to-spin conversion. This spin current exerts a torque on the magnetic moment **M**, which we call the OT. The torque efficiency ($\xi_{OT}$) of the OT is determined by not only the orbital Hall angle ($\theta_{OHE}$) of the LMs but also the spin-orbit conversion coefficient ($\eta_{L-S}$) of the FM layer. The $e$, $\hbar$, and $\sigma_{xx}$ are the charge of an electron, reduced Planck constant, and electrical conductivity of SHM and OHM, respectively.

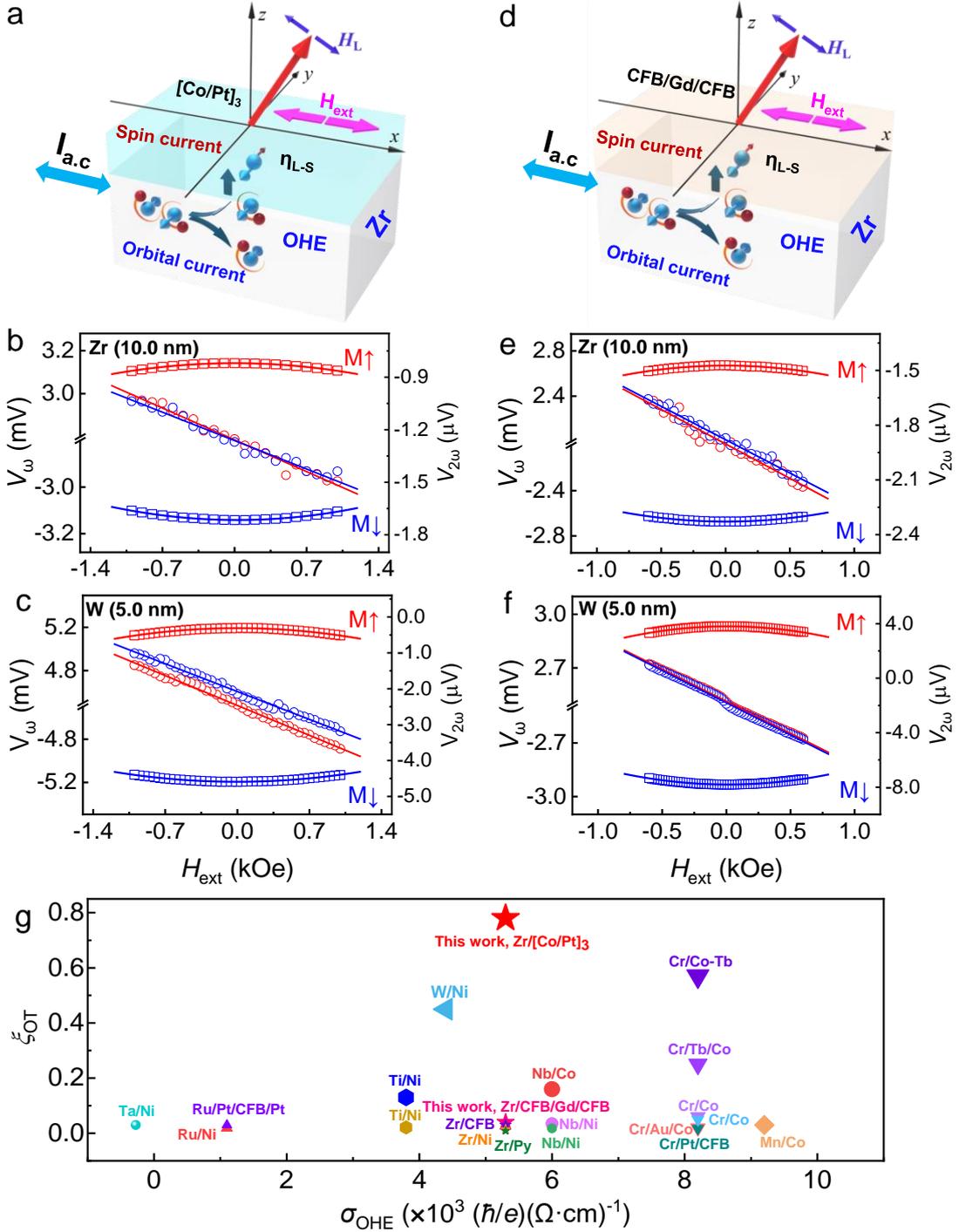

**Fig. 2. Orbital torque efficiency. a.** Schematic of the longitudinal orbital-torque induced effective field measurement for the [Co (0.3 nm)/Pt (0.7 nm)]$_3$ FM layer. **b, c.** The first harmonic voltage ($V_\omega$) vs. in-plane longitudinal external magnetic field ($H_{ext}$) and the second harmonic voltage ($V_{2\omega}$) vs. $H_{ext}$ for the Zr (10.0 nm)/[Co (0.3 nm)/Pt (0.7 nm)]$_3$ and W (5.0 nm)/[Co (0.3 nm)/Pt (0.7 nm)]$_3$ samples, respectively.

$\xi_{OT,DL}/\xi_{SOT,DL}$ is estimated to be ~ 0.78 and ~ 0.53 for the Zr OHM and W SHM systems, respectively. **d.** Schematic of the longitudinal orbital-torque induced effective field measurement for the CFB (0.8 nm)/Gd (1.2 nm)/CFB (1.1 nm) FM layer. **e, f.** The $V_\omega$ vs. $H_{ext}$ and the $V_{2\omega}$ vs. $H_{ext}$ for the Zr (10.0 nm)/CFB (0.8 nm)/Gd (1.2 nm)/CFB (1.1 nm) and W (5.0 nm)/CFB (0.8 nm)/Gd (1.2 nm)/CFB (1.1 nm) samples, respectively. $\xi_{OT,DL}/\xi_{SOT,DL}$ is estimated to be ~ 0.04 and ~ 0.35 for the Zr OHM and W SHM systems, respectively. **g.** The summary of the $\xi_{OT}$ as a function of $\sigma_{OHE}$ for the OHMs.

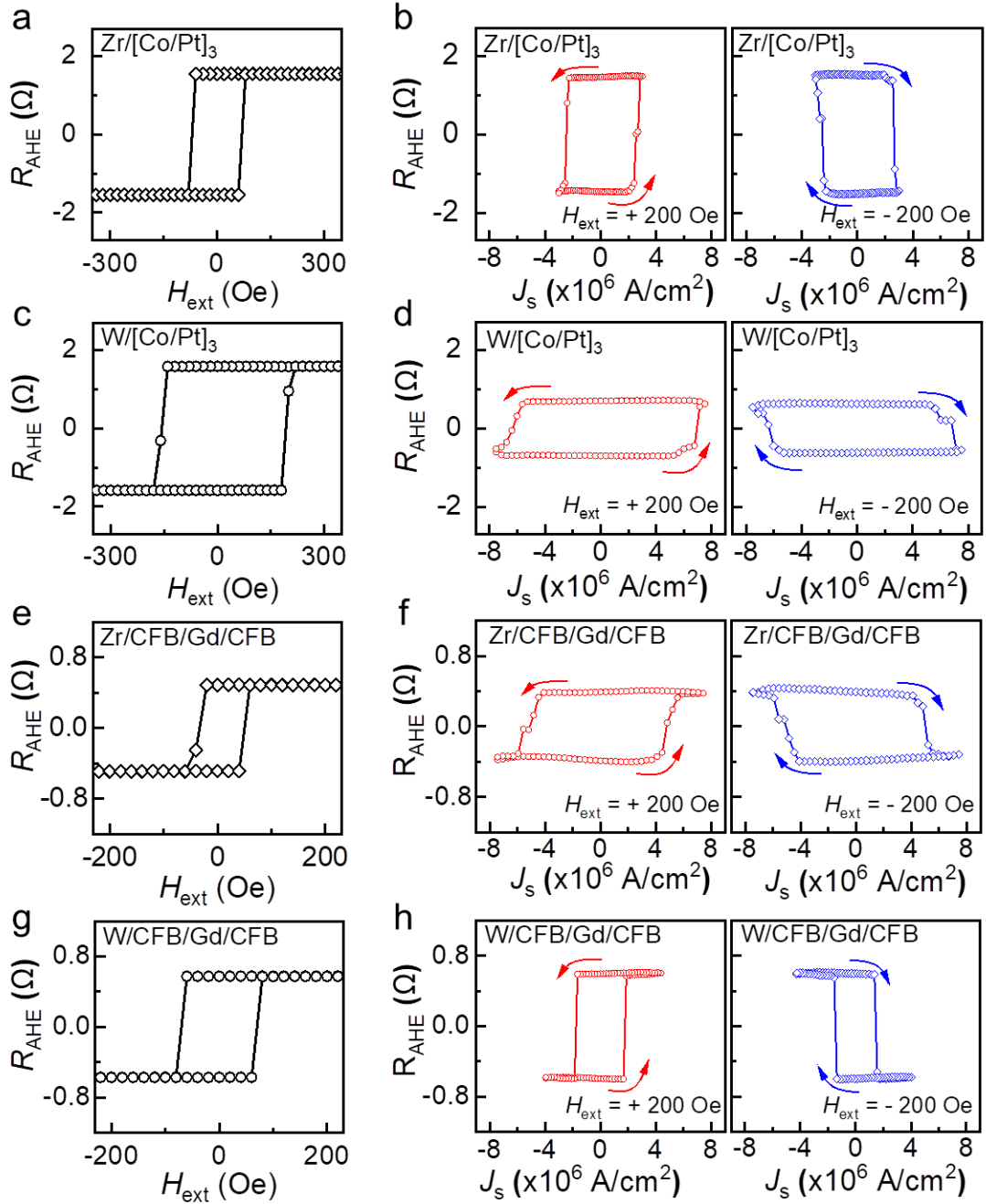

**Fig. 3. Orbital torque switching. a, c.** The AHE loops of Zr (10.0 nm)/[Co (0.3 nm)/Pt (0.7 nm)]$_3$ and W (5.0 nm )/[Co (0.3 nm)/Pt (0.7 nm)]$_3$ samples, respectively. **b, d.** Current-induced orbital torque switching for Zr (10.0 nm)/[Co (0.3 nm)/Pt (0.7 nm)]$_3$ and W (5.0 nm)/[Co (0.3 nm)/Pt (0.7 nm)]$_3$ samples with the fixed in-plane $H_{ext}$ = + 200 Oe and - 200 Oe, respectively. The $J_s$ of Z (10.0 nm)/[Co (0.3 nm)/Pt (0.7 nm)]$_3$ is as low as $J_s \sim 2.6 \times 10^6$ A/cm$^2$, which is lower than that of W (5.0 nm)/[Co

(0.3 nm)/Pt (0.7 nm)]$_3$ ($J_s$ ~ 6.6 × 10$^6$ A/cm$^2$). **e, g.** The AHE loops of the Zr (10.0 nm)/CFB (0.8 nm)/Gd (1.2 nm)/CFB (1.1 nm) and W (5.0 nm)/CFB (0.8 nm)/Gd (1.2 nm)/CFB (1.1 nm) samples, respectively. **f, h.** Current-induced orbital torque switching of the Zr (10.0 nm)/CFB (0.8 nm)/Gd (1.2 nm)/CFB (1.1 nm) and W (5.0 nm)/CFB (0.8 nm)/Gd (1.2 nm)/CFB (1.1 nm) samples with the fixed in-plane $H_{\text{ext}}$ = + 200 Oe and -200 Oe, respectively. The $J_s$ of Zr (10.0 nm)/CFB (0.8 nm)/Gd (1.2 nm)/CFB (1.1 nm) is ~ 4.9 × 10$^6$ A/cm$^2$, which is higher than that of W (5.0 nm)/CFB/Gd/CFB ($J_s$ ~ 1.6 × 10$^6$ A/cm$^2$).

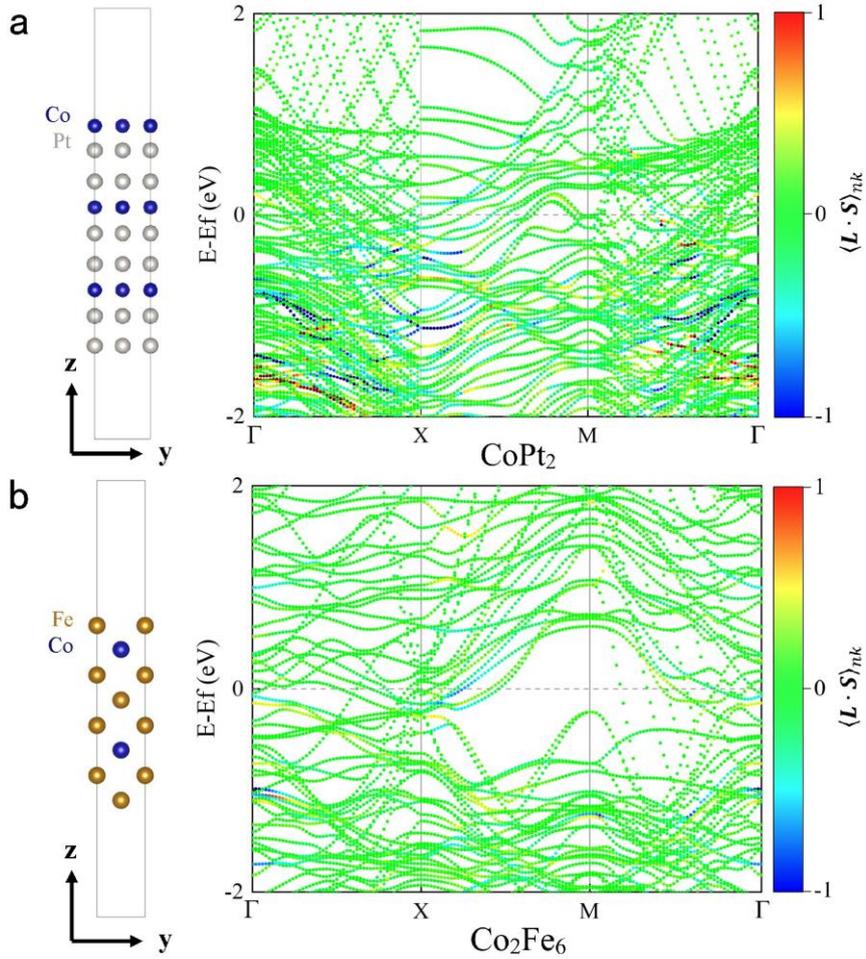

**Fig. 4. Theoretical calculation of spin-orbit correlation function** $\langle L \cdot S \rangle_{nk}$. **a.** The schematic of the designed $CoPt_2$ structure and the calculated result of the spin-orbit correlation function. **b.** The schematic of the designed $Co_2Fe_6$ structure and the calculated result of the spin-orbit correlation function. The color represents the correlation $\langle L \cdot S \rangle_{nk}$ for each eigenstate, in which red and blue denote strong positive and negative correlations, respectively.